# Semiconductor-to-metal transition in carbon-atom wires driven by $sp^2$ conjugated endgroups

*Alberto Milani [a], Matteo Tommasini [a], Valentino Barbieri [b], Andrea Lucotti [a], Valeria Russo[b], Franco Cataldo [c,d], Carlo S. Casari [b,*]*

[a] Department of Chemistry, Materials and Chem. Eng. 'G. Natta', Politecnico di Milano Piazza Leonardo da Vinci 32, I-20133 Milano, Italy

[b] Department of Energy, Politecnico di Milano via Ponzio 34/3, I-20133 Milano, Italy

[c] Actinium Chemical Research Institute, Via Casilina 1626A, 00133, Rome, Italy

[d]Università degli Studi della Tuscia, Dipartimento di Scienze Ecologiche e Biologiche, Viterbo, Italy

ABSTRACT: Novel bis(biphenyl)-capped polyynes have been synthesized to investigate the modulation of the electronic and optical properties of sp-hybridized carbon-atom wires (CAWs) capped with π-conjugated $sp^2$ endgroups. Raman and Surface Enhanced Raman spectroscopy (SERS) investigation of these systems and Density Functional Theory (DFT) calculations reveal structural changes from polyyne-like with alternating single-triple bonds towards cumulene-like with more equalized bonds as a consequence of the charge transfer occurring when wires interact with metallic nanoparticles. While polyynes have semiconducting electronic properties, a more equalized system tends to a cumulene-like structure characterized by a nearly metallic behavior. The possibility to drive a semiconductor-to-metal transition has been investigated by systematic DFT calculations on a series of CAWs capped with different conjugated endgroups revealing that the modulation of the structural, electronic and vibrational properties of the sp-carbon chain towards the metallic wire cannot be simply obtained by using extended π-conjugated $sp^2$ carbon endgroups, but require a suitable chemical design of the endgroup and control of charge transfer. These results provide useful guidelines for the design of novel sp-$sp^2$ hybrid carbon nanosystems with tunable properties, where graphene-like and polyyne-like domains are closely interconnected. The capability to tune the final electronic or optical response of the material makes these hybrid sp-$sp^2$ systems appealing for a future all-carbon-based science and technology.

Fullerenes, nanotubes and graphene represent a chief example of the versatility of carbon in producing nanostructures with different properties and dimensionality (zero, quasi-one and two, respectively). Such systems are mainly based on $sp^2$ hybridization, only one of the three different possibilities (sp, $sp^2$ and $sp^3$) available for carbon, two of which can also form stable allotropes occurring in nature (*i.e.*, $sp^2$ and $sp^3$ for graphite and diamond, respectively).[1] The long quest for the still lacking "third carbon allotrope" based on sp-hybridization has led to the development of many interesting nanoscale/molecular systems in the form of carbon-atom wires (CAWs).[2-5] CAWs are linear systems with finite length trying to approach the ideal infinite wire (*i.e.*, carbyne) representing the ultimate 1-D carbon system with intriguing properties, as evidenced by theoretical predictions [6-8]. The ideal carbyne can be in two different configurations: semiconducting wire with alternate single-triple bonds (*i.e.,* polyyne) and metallic wire with all double bonds (*i.e.*, cumulene). As finite systems, CAWs display structural electronic and optical properties strongly dependent on the length (*i.e.*, number of carbon atoms) and the type of functional termination (*i.e.*, endgroup) and are thus attracting for both fundamental and applied science.[2] Sp-hybridization naturally leads to one-dimensional model systems suitable to investigate quantum effects such as ballistic conductance, spin dependent electron transport and Peierls' distortion.[8-11] In a sp-carbon wire π-conjugation is maximized along one direction leading to outstanding properties, as predicted by theory. As a result in CAWs mechanical strength, thermal conductivity, electron mobility and nonlinear optical properties can reach unprecedented high values, outperforming many other materials so far investigated.[7,12-14] In addition, CAWs possess the tunability of properties enabling to have the insulator-semiconducting-metallic functionality in the same system, a feature that is still difficult to achieve or even lacking in other carbon nanostructures. Hence CAWs represent an appealing opportunity for developing functional nanostructures with tunable properties potentially exploitable for advanced applications.[2,15-17]

Today, CAWs can be synthesized in stable form with several different endgroups and have been investigated by diffraction and spectroscopy techniques. In particular, Raman spectroscopy is fundamental for the detection and structural investigation of CAWs, similarly to many other carbon-based systems.[18-22] Long CAWs up to 44 sp-carbon atoms have been reported by R. Tykwinsky and co-workers by rational chemical synthesis[23] in which the endgroups can be used both to stabilize the wire and for selecting polyyne vs. cumulene structure.[24] Recently, a wire exceeding 6000 atoms with a length of several hundreds of nm has been fabricated in the core of a double wall carbon nanotube acting as a protective cage.[25] Other interesting CAW-related systems have been produced by different chemical and physical approaches in liquids or in films.[26-30]

A system of particular interest is a CAW suspended between two graphene edges that can be fabricated with a top-down approach starting from graphene and removing atoms with the electron beam of a transmission electron microscope[31]. This achievement has stimulated a number of theoretical investigations focused on the peculiar electronic and transport properties of a sp-sp$^2$ hybrid carbon in which the graphene terminations naturally act as metallic electrodes of an atomic-scale device. On the experimental side, F. Banhart and co-workers reported the in situ study of the transport properties of a single wire bridging two graphene edges, outlining the occurrence of a strain-induced metal-to-semiconductor transition.[32] Very recently Sun et al. have reported the fabrication, with a surface science approach, and in situ STM observation of a sp-sp$^2$ hybrid 2-D system (*i.e.*, graphyne).[33] This peculiar hybrid system, according to theoretical predictions, should display mechanical, electronic and transport properties challenging those of graphene[34,35]. The development of nanoscale sp-sp$^2$ hybrid systems appears as a promising approach for a realistic exploitation of CAWs. To this aim a fundamental, yet still open issue, regards how and to what extent the properties of the sp-carbon wire can be effectively selected and modulated through the conjugation effect driven by sp$^2$ terminations. In this framework CAWs terminated by sp$^2$ carbon groups with increasing $\pi$-conjugation can be regarded as suitable model systems for investigating by a combined theoretical and experimental approach hybrid sp-sp$^2$ structures such as graphynes or graphene-wire-graphene systems.

Here we report an extensive analysis of the effect of different sp$^2$ endgroups on the structural, vibrational and electronic properties of CAWs. To this aim we have synthesized and characterized a new series of CAWs terminated by two biphenyl groups. Density Functional Theory (DFT) calculations together with Raman/SERS experiments unveil a semiconductor-to-metal transition induced by $\pi$-conjugation and charge transfer effects. This is discussed highlighting the role of aromatic endgroups of increasing size (i.e., phenyl, biphenyl, naphtyl and coronene) to show that the increase of the sp$^2$ conjugation alone is not enough to fully access to the tunability of properties. Additional strategies are here proposed as guidelines towards the control of functional properties of CAWs in a wide range of values from semiconductor to metal-like.

## RESULTS AND DISCUSSION

**Vibrational and structural characterization of bis(biphenyl)-capped polyynes.** The structure of bis(biphenyl)-capped polyynes (BPh[n]), resulting from DFT calculations, is presented in Figure 1 together with all the systems under investigation in this work. In particular we have considered hydrogen terminated polyynes (H[n]) and diphenyl-capped polyynes (Ph[n]), for comparison with simpler systems without or with minimal $sp^2$ conjugation, already considered in the past,[28,36-38] while dinaphtyl-capped (Naph[n])[39-41] and dicoronene-capped (Cor[n])[42] polyynes of increasing chain lengths, to assess the impact on molecular properties given by endgroups more extensively $sp^2$ π-conjugated than BPh[n]. In both Cor[n] and Naph[n] the molecules can adopt the anti or the syn configuration. We limited our DFT calculations to the syn configuration since it has been demonstrated that π-conjugation and related properties (structure, gap, Raman frequencies) are essentially the same in both configurations [41,42]. The results obtained for the above mentioned polyynes have been compared with those reported for long cumulenes (tBuPh(n) and Mes(n)) recently synthetized [24,43] and investigated by DFT calculations.[44]

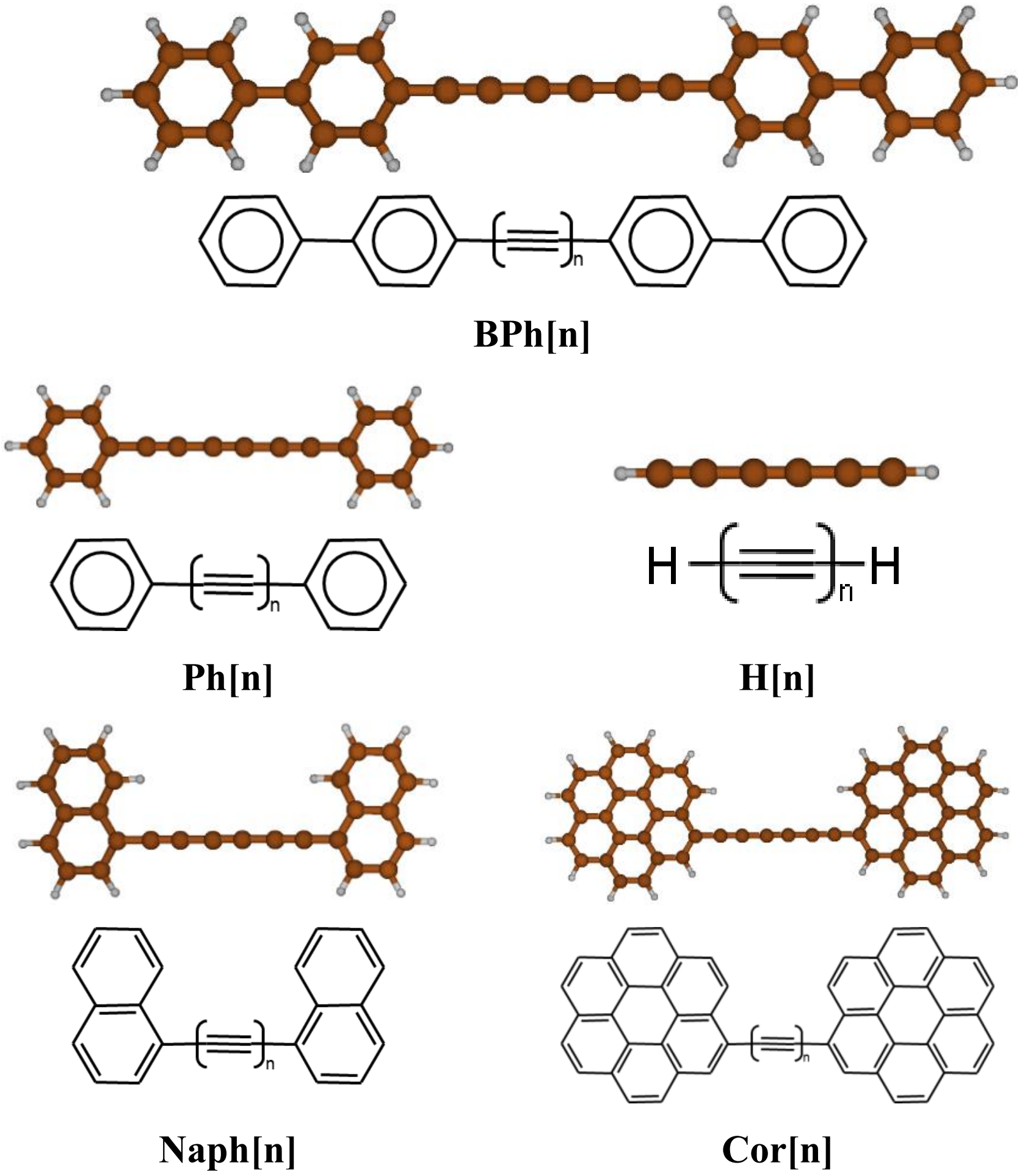

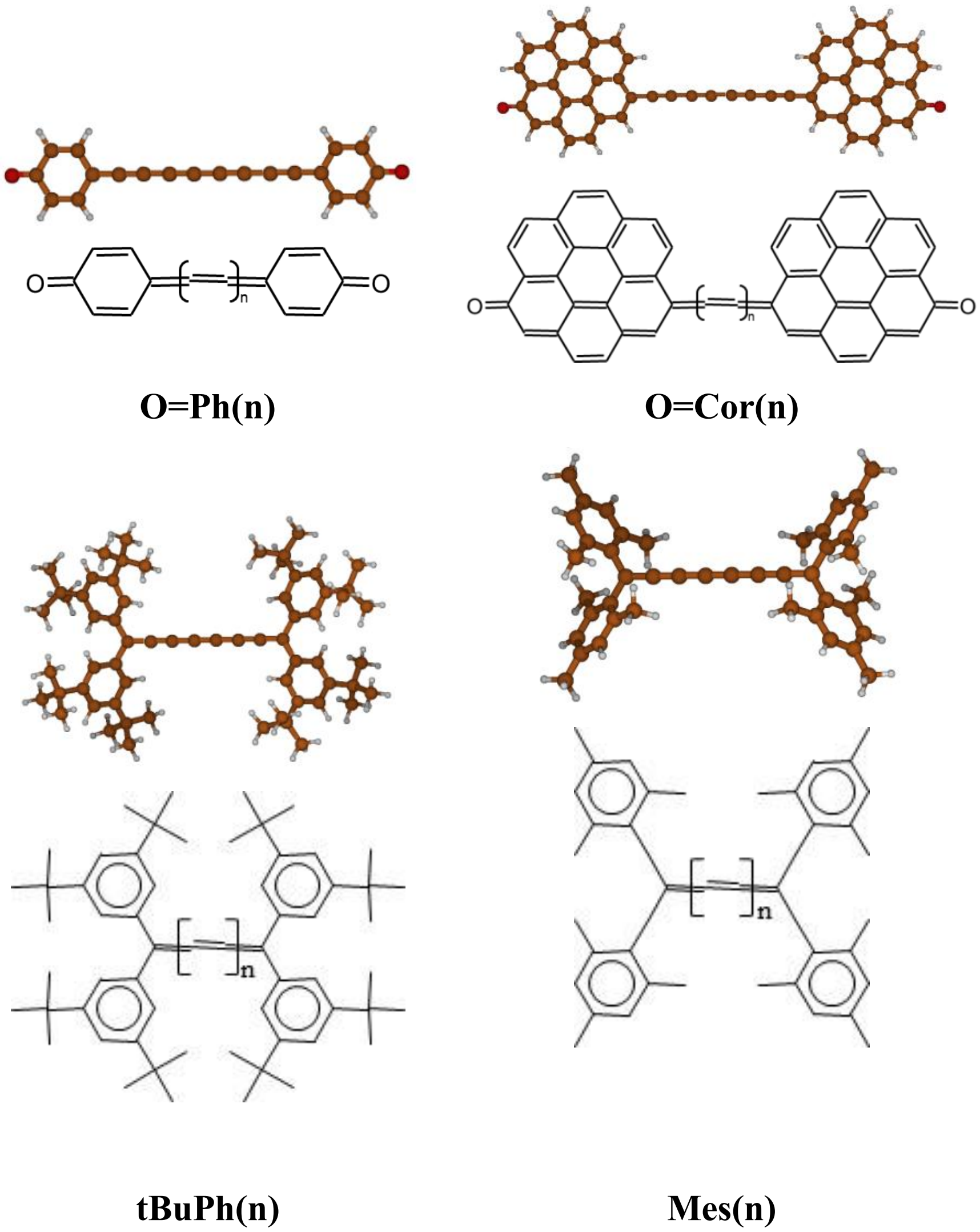

**Figure 1.** Sketches of the CAWs investigated by means of DFT calculations. The number n of the sp hybridized CC units in the molecules ranges from 2 to 6 (i.e. 4 to 12 carbon atoms in the sp wire).

The Raman spectrum of BPh[n] in solution shows well defined features in the 1950-2250 $cm^{-1}$ region which is a typical signature of sp-carbon (see Figure 2). The peak at 1600 $cm^{-1}$ is attributed to the ring-stretching modes of the phenyl endgroups. The sp-carbon features reveal the presence of wires with different lengths as confirmed by the size distribution obtained by HPLC showing lengths in the 4-12 atoms range. Following the results from DFT calculations and a procedure adopted in the similar case of Ph[n][28], each peak can be assigned to a specific wire length (see Figure 2). All of these peaks have a wavenumber above 2000 $cm^{-1}$, which is typical of short wires with polyyne-like structure. The data displayed in Figure 2 clearly evidence the red-shift of the most intense Raman line with increasing chain length. The associated wavenumber dispersion has been rationalized in the framework of the Effective Conjugation Coordinate (ECC) model, which was proposed for the interpretation of the Raman spectra of polyconjugated materials[45]. The intense ECC line observed in the Raman spectra is assigned to a collective CC stretching normal mode, which is described as an oscillation of the Bond Length Alternation (BLA), defined as the difference between the average values of sp-hybridized CC bond lengths of alternated longer and shorter bonds in CAWs. This mode red shifts as the chain length increases due to increasing π electron delocalization. Such behavior has been investigated in details for sp-hybridized linear carbon chains.[2,18,46,47]

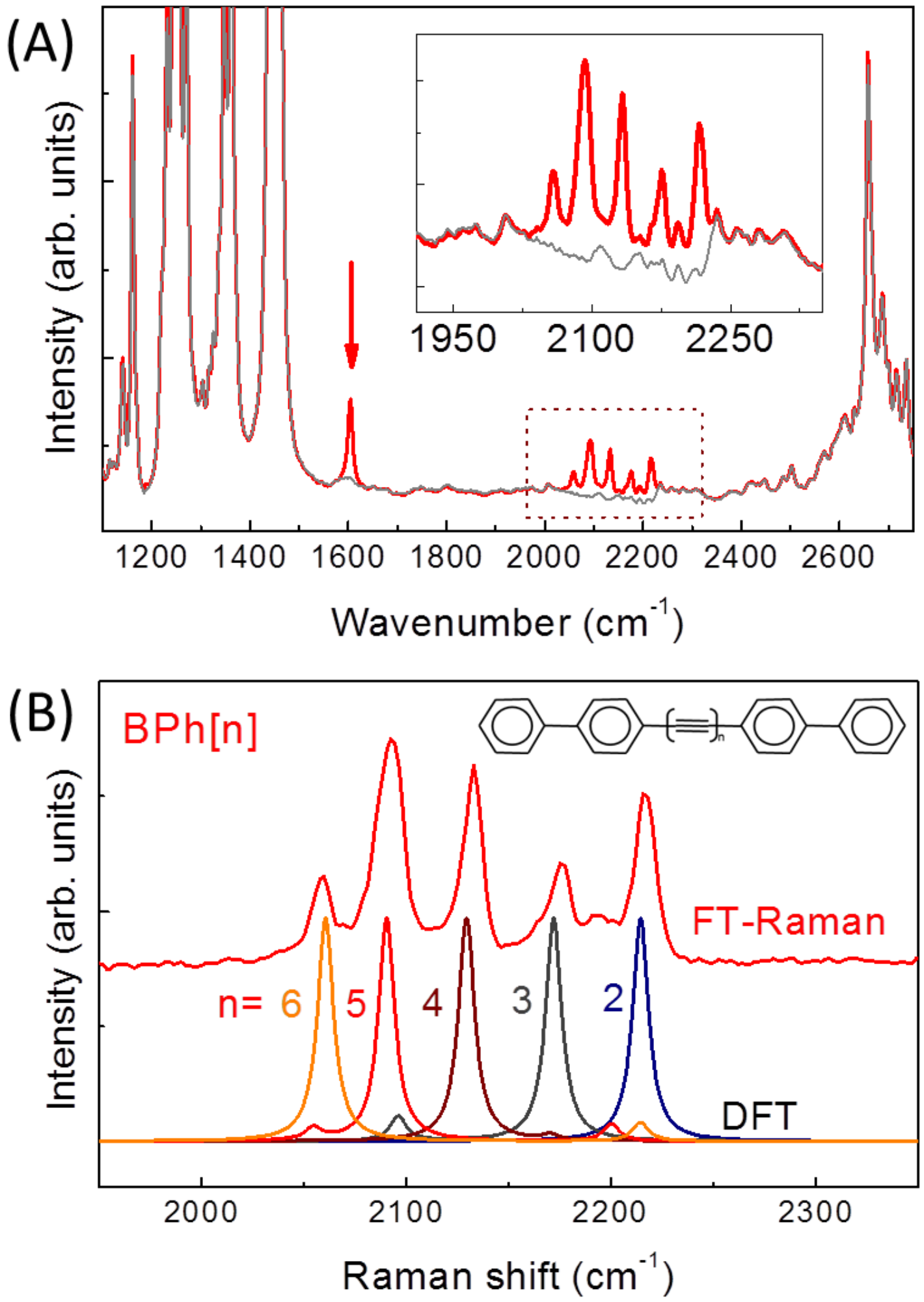

**Figure 2.** (A) FT-Raman spectrum of BPh[n] polyynes in decalin solution (1064 nm excitation). The background signal of the solvent is reported in grey for comparison. The inset provides a detailed view of the spectral region of interest for sp-carbon. (B) Identification of the contribution of different chain lengths in the spectrum of BPh[n] in solution according to DFT calculations of the Raman response.

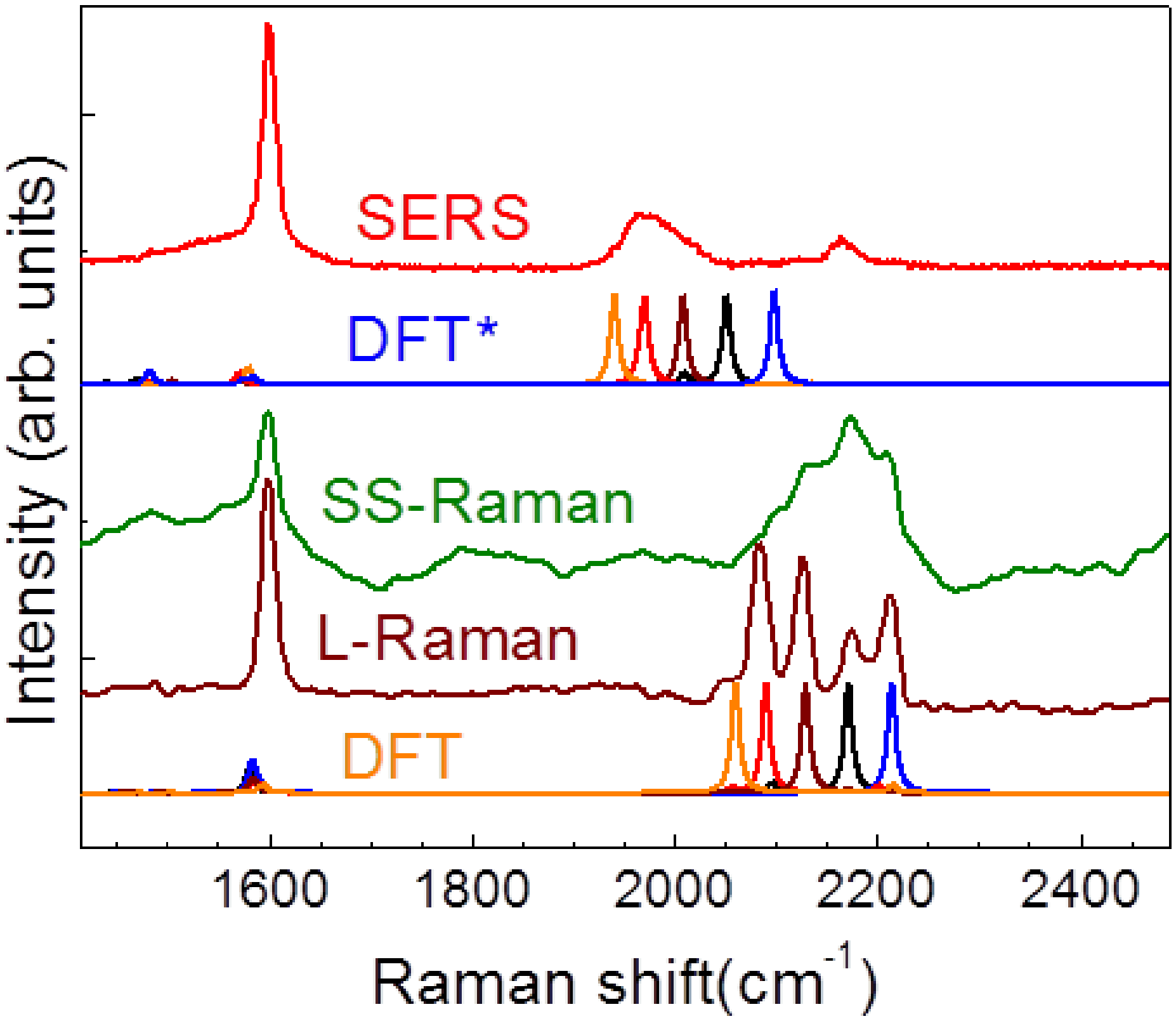

**Figure 3.** Experimental SERS spectrum at 514.5 nm and DFT computed Raman spectra for negatively charged (-1) model systems (DFT*). FT-Raman spectrum in liquid (L-Raman) at 1064 nm, solid state Raman (SS-Raman) at 514.5 nm and DFT computed Raman spectra for neutral wires (DFT) are also shown for comparison.

The trends observed in Raman spectra gain a particular relevance by considering the close link existing among the sp-CC stretching frequencies, the electronic gap and the structure of CAWs.[48] The increase in chain length implies a larger degree of π-electron delocalization which affects structural, electronic and vibrational properties. With increasing CAWs length, the BLA decreases and this behavior parallels the decrease of the HOMO-LUMO gap and the progressive red-shift of the frequency of CC stretching vibrations observed with Raman spectroscopy. The interplay existing among these physical quantities has been analyzed in many papers by means of experimental and theoretical approaches[2,18]. As expected, BPh[n] polyynes show a steady decrease of BLA with chain length. For instance, BLA changes from 0.1413 Å to 0.1170 Å moving from 4 to 12 carbon atoms (i.e., from BPh[2] to BPh[6]). This is correlated to a decrease of the HOMO-LUMO gap from 3.95 eV to 3.16 eV (see Supporting Information) and to the decrease of the ECC wavenumber experimentally observed by Raman spectroscopy (see Fig. 2 and Supporting Information).

The conformation of the aryl endgroups of BPh[n] polyynes can have a further effect in modulating π-electron conjugation which results to be maximized when all the aryl groups lie in the same plane, as observed in the case of Ph[n][28]. However, due to the steric hindrance of the aryl hydrogens such a conformation is not stable. Instead, DFT calculations predict an equilibrium dihedral angle of about 36° between the planes formed by nearby aryl groups.

SERS spectrum of BPh[n] polyynes shows remarkable differences in comparison with normal Raman (see Figure 3). Besides the peak at 1600 $cm^{-1}$ (stretching mode of the aromatic aryl units), which is just enhanced in relative intensity (here used to normalize the intensity in the spectra of figure 3), the spectral region assigned to sp-CC stretching (ECC lines) is characterized by the appearance of a new broad feature below 2000 $cm^{-1}$. At higher wavenumber the SERS spectrum is less structured than the Raman spectrum and a main peak is observed at about 2160 $cm^{-1}$. The pattern observed in SERS is peculiar, with the broad band at about 1950-2000 $cm^{-1}$ occurring at much lower wavenumbers than any of the sp-CC stretching Raman lines. A reminiscence of the features observed in liquid is still present as evidenced from the comparison of normal Raman spectrum on a film (i.e. sample solution drop casted and dried on substrate). The Raman/SERS data collected on BPh[n] polyynes (Fig. 3) parallel the behavior of Ph[n] polyynes which was investigated in the past[28]. This behavior was explained by a charge transfer occurring between the metal nanoparticles (Ag or Au) and the adsorbed Ph[n] polyynes. As indicated by DFT calculations, the charge transfer causes a decrease of BLA and the consistent occurrence of Raman signals significantly red-shifted with respect to that of the neutral Ph[n] polyynes[28]. To support this interpretation in the case of BPh[n] polyynes, we carried out DFT calculations also for positively and negatively charged model systems. As expected, in the charged species we obtain an equalization of the bond lengths (BLA decrease), moving towards a polyyne-to-cumulene transition induced by charge transfer. The BLA values of charged molecules are systematically and significantly smaller than those of neutral molecules. In BPh[4] the BLA has a reduction of 30% from 0.125 Å to 0.0875 Å while passing from the neutral to the anion form. On the other hand, the trends of BLA vs. chain length in charged BPh[n] are similar to those found in the neutral state (see Table S1). Furthermore, DFT calculations on charged species reveal that the sp-CC stretching Raman line corresponds to the position of the broad band observed in the SERS spectra (see Table S2). This proves that charge transfer effects significantly affect the structure and vibrational properties of BPh[n] polyynes.

In accordance with the behavior of Ph[n] polyynes, the calculation of the ionization potentials (IP) and electron affinities (EA) of BPh[n] polyynes, compared with those of Ag, reveals that the charge transfer occurs from the metal nanoparticle to the polyyne. This is proved by the calculation of the work required for the charge transfer, i.e. $E_{ion}$ [$E_{ion}$ = IP − EA], considering the two possible scenarios (Ag+/polyyne- or Ag-/polyyne+), as described in Ref. [28]. The values of $E_{ion}$, IP and EA are reported in SI. It is possible to verify that $E_{ion}$ is lower in BPh[n] than in Ph[n] polyynes. This is consistent with the slightly larger π-conjugation of BPh[n] compared with Ph[n] at a given chain length.

**Tuning the structure and electronic properties by selection of $sp^2$ endgroups.** The previous analysis reveals that, compared with Ph[n] polyynes, the presence of slightly more conjugated endgroups in BPh[n] affects the structure of CAWs, and causes a slight decrease of the BLA and HOMO-LUMO gap, together with a red-shift of the ECC Raman lines.

These findings open up interesting perspectives for potential exploitation of CAWs, also in relation with their coupling with graphene or graphitic domains. Hence, a fundamental point is how and to what extent it is possible to tune the structural, electronic and vibrational properties of CAWs by means of a proper choice of the endgroups. To this aim, we compare the properties of BPh[n] with those of other experimentally available polyynes/cumulenes which have been synthetized and/or investigated in the recent past. The list of the polyynes here considered includes Ph[n][28] (closely related to BPh[n]), Naph[n][40], Cor[n][42] and hydrogen capped H[n][49] (taken as well-established reference system). Among cumulenes we include the recently synthetized tBuPh(n)[24,43] and two oxidized variants of the polyynes mentioned above, namely O=Pn(n) and O=Cor[n] (see Fig. 1). The comparison between BPh[n] and Ph[n] polyynes allows investigating the effects of increased π-conjugation caused by the replacement of phenyl caps with biphenyl caps. The comparison of the FT-Raman spectra of these two series of polyynes, shown in Figure 4 reveals that sp-CC stretching lines of BPh[n] are systematically red-shifted with respect to Ph[n] polyynes of corresponding chain length (n ranging from 2 to 6 corresponding to wires with 4 to 12 sp carbon atoms).

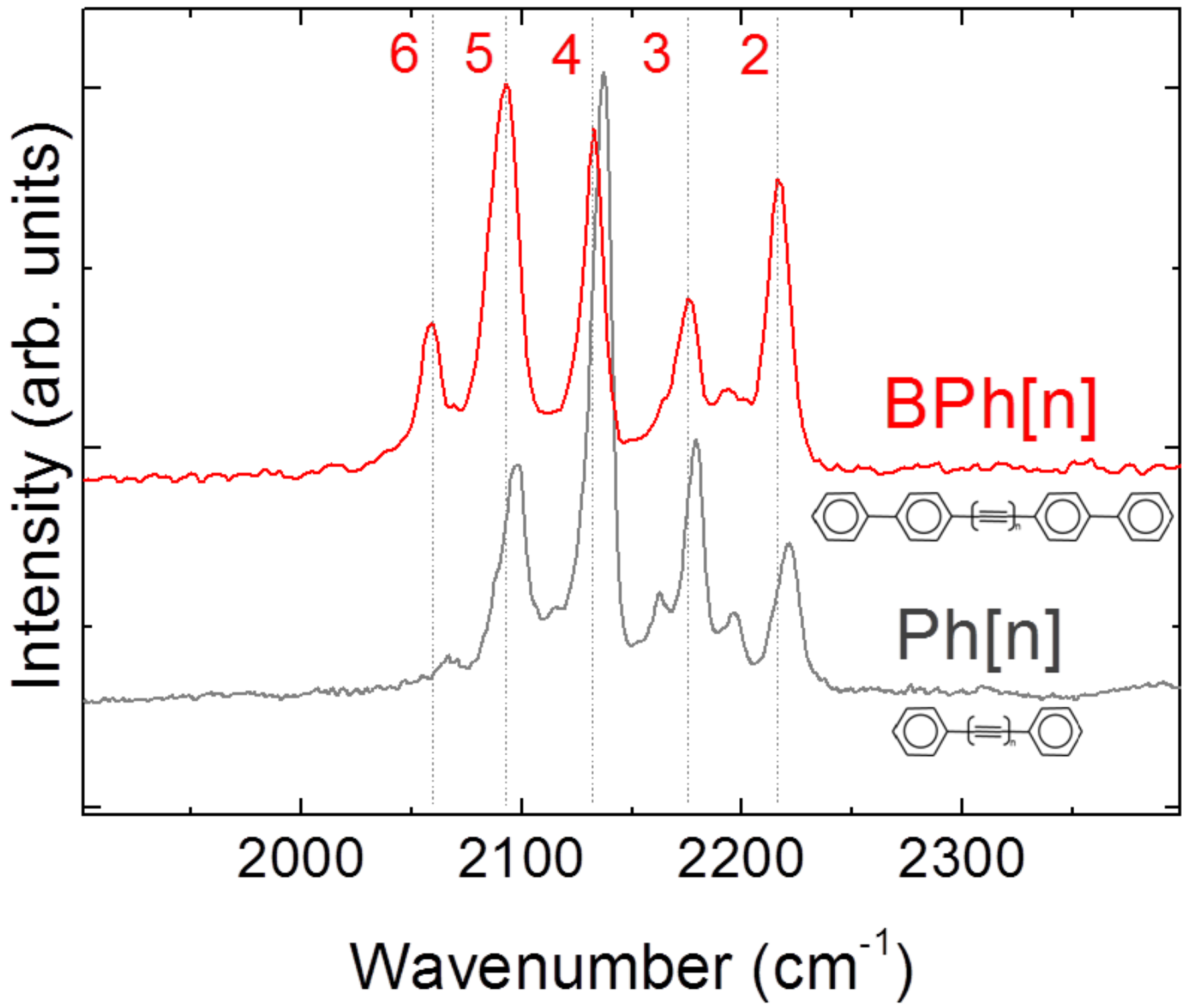

**Figure 4.** Experimental FT-Raman spectra of BPh[n] molecules compared with the FT-Raman spectra of parent Ph[n] polyynes[28].

This is expected due to increased π-conjugation and it is also supported by DFT calculations (Table S3 of the Supporting Information). Furthermore, the behavior observed with vibrational spectroscopy is consistent with the fact that the BLA values of BPh[n] are lower than those of Ph[n] (Table S1 of the Supporting Information). This proves the close relationship existing between the structure and vibrational properties of CAWs and the great sensitivity of Raman spectroscopy, even to minor changes in π-conjugation.

Interestingly, the BPh[n] and Ph[n] samples display different relative Raman intensities (see Fig. 4). In particular BPh[n] show larger relative intensity for the peaks associated to the longer chains than Ph[n], which qualitatively indicates a larger amount of longer chains in the BPh[n] sample and suggests that BPh[n] are more stable with respect to Ph[n] wires. In fact, when the sp-hybridized skeleton interacts with neighboring wires, crosslink reactions may lead to phase change towards $sp^2$ domains. This tendency which is naturally larger for increasing chain lengths can be prevented by selecting bulky endgroups, which are able to keep CAWs at distance, thus protecting the sp backbone against mutual interactions and crosslinking[50]. Hence, the larger amount of longer chains in the BPh[n] sample can be explained by the presence of bulkier endgroups with respect to Ph[n]. In fact, compared to phenyl caps, biphenyl caps may generate a larger steric hindrance also as a result of their non-planar conformation.

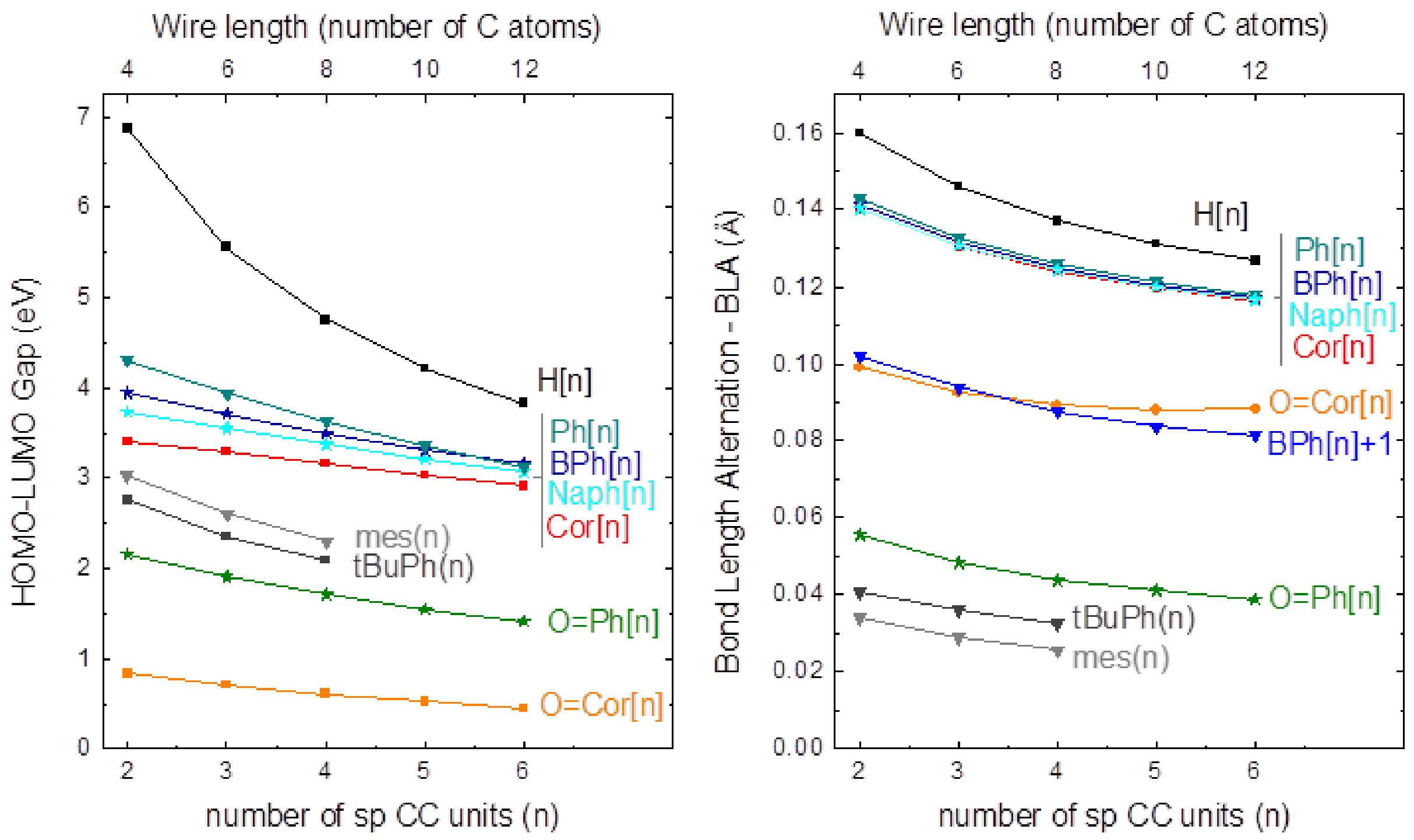

**Figure 5.** Plot of the DFT computed values of BLA (defined as the difference between the average values of sp-hybridized CC bond lengths of alternated longer and shorter bonds in CAWs) in A and HOMO-LUMO gap in eV of the CAWs reported in Figure 1. BLA has been determined also for BPh[n] anionic and cationic species (see text for details)

In Figure 5, we extended the computational analysis to the other polyynes/cumulenes reported in Figure 1. Starting with H[n], polyynes with non-conjugated caps (here used as a reference), it is evident that the terminal CH bonds (single) impose a triple bond on the first CC bond of the chains thus determining the overall polyyne-like structure of the wire. This is evident from the large values of BLA and electronic gap. Increasing the H[n] wire length can reduce both the electronic gap and the BLA even if in this case the reduction is somehow limited and does not affect the structure and the electronic behavior of the system to a wide extent.

In comparison with H[n], the adoption of a phenyl endgroup (Ph[n]) is less effective in imposing a triple bond on the first CC bond of the wire, due to the conjugated character of the $sp^2$ carbons of phenyl, resulting in a lower BLA (about 10% reduction of the BLA, from 0.16 to 0.14 A for the shortest wire) and a significantly lower HOMO-LUMO gap (37% of reduction from 6.9 eV to 4.3 eV). This result suggests that π-conjugated endgroups ($sp^2$-hybridized) can effectively communicate with the sp-carbon wire[28,47]. In fact, comparing BPh[n] with Ph[n] polyynes, the slightly larger π-conjugation of biphenyls affects the band gap, the BLA and the sp-CC stretching frequency, as mentioned above. However, these changes are quite small and almost negligible in the case of BLA (see Table S1 and Figure 5). The non-planar conformation of the biphenyl group and its torsional flexibility certainly has a role in such a behavior. Thus we move to the case of Naph[n] polyynes where the two rings inside each endgroup are condensed and are forced to assume a planar geometry and a larger π-conjugation. Intriguingly, the effect on BLA is very small also in this case and a similar situation is found even considering more extensively π-conjugated coronene endgroups (Cor[n] polyynes). This indicates that π-electron delocalization between the $sp^2$ and sp regions of the molecules is not so effective in the modulation of the properties of the wires. Considering the HOMO-LUMO gap a somewhat larger effect is found, particularly in shorter polyynes which are also expected to be more sensitive to increasing π-conjugation. However, the modulation of the gap is restricted within a quite small range, similarly to the case of BLA (see Figure 5). Therefore, these results seem to rule out the possibility to significantly affect the structural and electronic properties of polyynes by choosing proper $sp^2$ carbon based endgroups.

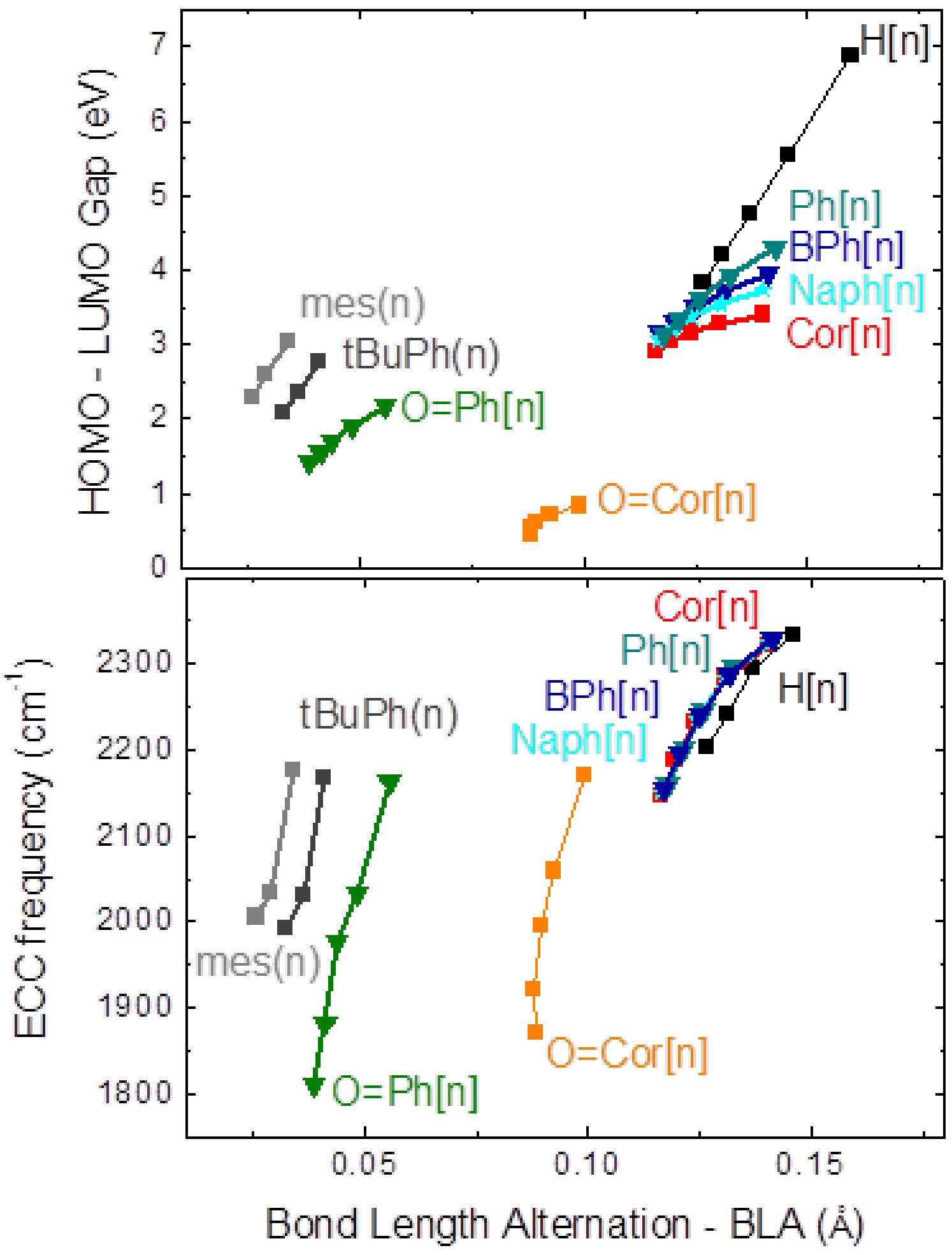

**Figure 6.** Left: Plot of the DFT computed HOMO-LUMO gap (eV) as a function of BLA (A); right: plot of the DFT computed ECC wavenumber ($cm^{-1}$) as a function of BLA (A)

A rather different situation is found in long cumulenes tBuPh(n), which have been recently synthesized [24,43] and fully characterized through vibrational spectroscopy and DFT calculations[2,47]. The plots reported in Figs. 5 and 6 and data in Tables S1-S3 (Supporting Information), reveal very low BLA values in tBuPh(n) cumulenes, corresponding to a low electronic gap. The choice of tBuPh endgroups effectively tunes the structure and the properties of the sp carbon wire towards a metallic system. The molecular structure of tBuPh(n) cumulenes imposes that the terminal C atom of the wire forms two single bonds with the aryl endgroups. This implies that the first CC bond of the sp-chain has a double bond character which henceforth extends to the whole sp-chain generating a cumulene-like structure. Such behavior is similar to the case of cumulenic vinylidene-capped CAWs which show a comparable low value of BLA (0.0306 Å[51] with respect to 0.0361 Å for tBuPh(6) and 0.02883 Å for Mes(6)).

This result demonstrates that chemical strategies other than the choice of more conjugated endgroups should also be taken into consideration for tuning the properties of CAWs. In fact, by introducing π-conjugated endgroups, the charge delocalization to the wire is effective but small and seems to level-off and saturate for increasing size (i.e., conjugation path length) of the endgroups. On the other hand, if the endgroups are chosen such that they impart a double bond character to the first CC bond of CAWs, then a more equalized structure (towards cumulene) is obtained as a result of a sort of domino effect where the first C=C bond propagates its character all along the sp-chain.

To exploit the potentialities of mixed graphene/carbyne-like hybrid systems, our findings suggest that other strategies could also be adopted to get tunable conductive (i.e., metallic) CAWs. A further possibility in mixed $sp^2$/sp systems is the selective oxidation of the terminal endgroup. This oxidation can be used to suitably induce a quinoid structure on the $sp^2$ endgroup which would then affect the conjugation pattern and impose a cumulene-like (nearly-metallic) structure on the sp-chain. To prove this concept we carried out a calculation on the Ph[4] polyyne where one of the CH bonds has been substituted with a C=O bond. Such a bond promotes indeed a peculiar localization of the C=C bond on the endgroups which in turns induces a double bond character on the first CC bond of the chain, generating in cascade a possible cumulenic structure. As expected, the resulting O=Ph(4) molecule shows a very low BLA value, which is similar to that of long cumulenes (see Fig. 5 and Table S1). For this reason we name the system as O=Ph(4), adopting the notation of round parenthesis referring to its cumulene-like structure (see Fig. 1). Interestingly, the HOMO-LUMO gap and the sp-CC stretching wavenumber of O=Ph(4) are even smaller with respect to tBuPh(n) cumulenes. The same calculation has been also carried out for an oxidized Cor[n] polyyne, leading to O=Cor(n). In this case BLA is lower than the unsubstituted molecule Cor[n] but it is larger than that of long (tBuPh(n)) cumulenes while the HOMO-LUMO gap is much smaller.

The previous discussion considers neutral molecules. As already commented above, a further possibility to tune CAW properties may be the use of charge transfer effects to significantly lower the BLA and induce a transition toward metallic species. In the charged CAWs here considered the CC bonds tend to equalize, both for cations and anions (Fig. 5). The trend of the ECC wavenumber also supports the significant increase of π-conjugation found in charged species (see Supporting Information). Therefore, one promising approach to tune the properties of mixed sp-$sp^2$ carbon systems could be a controlled modulation of the charge-transfer between the systems and a proper substrate/nanostructure. The advantage of this approach is also the possibility to induce the metallic property in a reversible way by controlling the charge transfer without the need of complex synthesis procedure to select the endgroup.

Finally, for all the molecules considered here, we report the plot of ECC wavenumber vs. BLA in Fig. 6, which clearly displays the correlation existing between the structure and the vibrational properties of CAWs. The reported trends are in good agreement with the results of the infinite chain models predicting a monotonic increase of the ECC wavenumber for increasing BLA. In fact in Ref. [48] a similar plot was reported, based on a simple theoretical model (Hückel theory) worked out for the ideal infinite chain (i.e., carbyne). Furthermore, for all the considered systems the ECC behavior as a function of BLA shifts towards higher BLA values for CAWs with higher HOMO-LUMO gap, in agreement with the theoretical model presented in ref. [48]. Thus, the ECC vs. BLA plot constitutes a very effective summary describing the correlation existing between structural, vibrational and electronic properties (i.e., larger BLA ←→ larger HOMO-LUMO gap ←→ larger ECC wavenumber). The data collected in this work clearly show that the modulation of the physical properties of CAWs can be driven by endgroups over a rather wide range.

In conclusion we underline that: i) by using proper chemical substitutions affecting the conjugation pattern or by exploiting charge transfer effects, it is possible to obtain mixed sp-sp$^2$ carbon systems with tunable semiconductive/metallic behavior and tunable optical gap over a wide range; ii) a computational approach based on DFT calculations combined with experimental investigations based on vibrational spectroscopy display promising and powerful molecular engineering capabilities. All these findings call for further investigations aimed at exploring a wider range of chemical substitutions and assessing the best preparation strategies for selected CAW properties.

## CONCLUSIONS

Carbon-atom wires are currently attracting as experimentally available nano/molecular systems with appealing properties approaching those of the ideal carbyne. The major difficulties for the full exploitation of these systems regard stability and the effective tunability of properties opened by the control of length and by the choice of suitable endgroup. In finite systems the endgroup plays a relevant role in determining the structural organization of the wire and hence in defining the electronic and optical properties. Recent achievements have demonstrated that $sp^2$ conjugated endgroups have the advantage of improving stability and enhancing the overall conjugation of the system, even though it is still not completely clear how to drive properties to have nearly zero gap and metallic features (i.e., cumulene-like systems).

We have here shown that $sp^2$-based endgroups are effective in reducing the gap and the BLA towards cumulene-like metallic structures even though the effect rapidly saturates with increasing the size of the endgroup (i.e., number of aromatic units). Our investigations have outlined that the insertion of C=O units in the endgroup can further enhance π conjugation, indicating the importance of selecting precise chemical and functional properties of the endgroup. This suggests graphene oxide (GO) as a termination of wires to drive the functional properties of the system, e.g. in a GO-CAW-GO nanoscale device. In addition, the control of charge transfer, *e.g.* upon interaction with metal nanoparticles is proposed as an additional route towards cumulene-like structures. Hence, combining these strategies (i.e., proper choice of the endgroup and charge transfer) could be an effective strategy to control CAWs towards the metallic behavior. Our findings can be used as general guidelines for developing functional wires with engineered properties spanning from semiconducting to nearly-metallic systems with great interest for science and with potential applications in advanced nanotechnology systems and devices.

## MATERIALS AND METHODS

**Materials.** Diiodoacetylene prepared as described in ref. [52] was dissolved in decalin and stirred with copper biphenylacetylide (details of the preparation method will be published elsewhere). After purification the solution was analyzed by HPLC C8 column showing the presence of 1,4-bis(biphenyl)-butadiyne, 1,6-bis(biphenyl)-hexatriyne, 1,8-bis(biphenyl)-octatetriyne, 1,10-bis(biphenyl)-decapentiyne, 1,12-bis(biphenyl)-dodecahexiyne (named hereafter as BPh[2], BPh[3], BPh[4], BPh[5] and BPh[6], respectively). The relative concentration was determined from the electronic absorption spectra measured with the diode array detector. For BPh[2] the relative concentration in the mixture was 77.1% by mol, 18.4% for BPh[3], 2.6% for BPh[4], 1.7% for BPh[5] and 0.2% for BPh[6].

**Spectroscopic Measurements.** Micro-Raman spectra were acquired with a Renishaw Invia equipped with a 514.5 nm Ar+ laser line. Measurements were carried out after drop-casting and drying the sample solution (in decalin solvent) on silicon or glass substrates. For each solid sample Raman spectra were acquired at several different spots to check the homogeneity of the sample. Unless otherwise stated, Raman spectra were collected focusing through a 50x microscope objective at a laser irradiation power of 1 mW.

FT-Raman spectra were acquired with the Nicolet NXR 9650 instrument equipped with a Nd-VO4 laser (providing a 1064 nm excitation) and an InGaAs detector. Spectra were recorded directly on polyyne solutions placed within quartz NMR tubes. Spectra of pure solvents were also recorded as a reference background. Since the minimum spot size (~50 μm) is much larger than in micro-Raman measurements, a higher laser power is required to reach a comparable power density at the sample and backscatter a significant Raman signal. Therefore, measurements on BPh[n] were taken with an incident power of 2.5 W averaging over 1024 acquisitions (each lasting ~2 seconds).

SERS spectra were acquired by drop-casting the sample solution on SERS-active substrates made by silver nanoislands evaporated on silicon or glass. The equivalent thickness of silver (measured by a quartz microbalance) was finely adjusted to tune the plasmon resonance peak (measured by UV-Vis absorption spectroscopy) as close as possible to the laser excitation used in the micro-Raman instrument. The SERS-active substrates were carefully cleaned before use in ultrasonic bath (isopropyl alcohol, 5 minutes). This allows excluding any spurious background signal over the spectral region of interest for sp-hybridized carbon species (1800-2200 $cm^{-1}$).

**Computational Methods.** The prediction of the structural, vibrational and electronic properties of all the systems here investigated has been carried out in the framework of density functional theory (DFT). We adopted the hybrid PBE0 exchange-correlation functional[53] and the cc-pVTZ basis set. All calculations were carried out with the Gaussian09 quantum chemistry code[54]. For each molecule we have carried out geometry optimization and the calculation of the IR and Raman spectra (for the latter we adopted off-resonance condition). It was reported[55,56] that the PBE0/cc-pVTZ method is well suited for reliably simulating the Raman response of CAWs. A good agreement with experimental data is expected, as reported in a previous work on related systems, i.e., diphenyl-capped polyynes (Ph[n])[28]. However, since current DFT methods are known to overestimate π-electron conjugation and predict an exaggerated frequency softening with increasing chain length, suitable scaling procedures are needed[55,57] when comparing in details simulated spectra with experimental spectra. In the case of BPh[n], due to the similarity with Ph[n] polyynes investigated in the past, we adopted the same scaling procedure[28]. This consists in introducing a set of chain length-dependent scaling factors which are determined on the basis of the corresponding lines observed in the experimental Raman spectra. These scaling factors are 0.9503, 0.9487, 0.9508, 0.9522, and 0.9565 for n = 2, n = 3, n = 4, n = 5 and n = 6, respectively and have been used also for BPh[n] molecules. Further explanations about the rationale of this method are given in Ref. [28]. With the same theoretical approach, to help the interpretation of observed SERS spectra on the basis of charge-transfer effects we have also carried out calculations for BPh[n] in cationic and anionic states (±1 charge).

## ASSOCIATED CONTENT

### Supporting Information

Tables with DFT computed numerical values of BLA, HOMO-LUMO gap and ECC wavenumbers of the investigated models; description of the procedure adopted to investigate charge-transfer effects; plot of the ECC wavenumber as a function of chain length.

## AUTHOR INFORMATION

### Corresponding author

E-mail: carlo.casari@polimi.it (Carlo S. Casari)

## REFERENCES

(1) Hirsch, A.; The era of carbon allotropes. *Nat. Mater.* **2010**, 9, 868-871.

(2) Casari, C.S.; Tommasini, M.; Tykwinski, R.R.; Milani, A. Carbon-atom wires: 1-D systems with tunable properties. *Nanoscale* **2016**, 8, 4414-4435.

(3) Tarakeshwar, P.; Buseck, P.R.; Kroto, H.W. Pseudocarbynes: charge-stabilized carbon chains. *J. Phys. Chem. Lett*. **2016**, 7, 1675−1681

(4) Banhart, F. Chains of carbon atomo: a vision or a new nanomaterial. *Beilstein J. Nanotechnol*., **2015,** 6, 559-569

(5) Baughmann, R.H. Dangerously seeking linear carbon. *Science* **2006**, 312, 1009-1110

(6) Artyukhov, V.I.; Liu, M.; Yakobson, B.I. Mechanically induced metal insulator tranition in carbine. *Nano Letters* **2014**, 14, 4224–4229.

(7) Liu, M.J; Artyukhov, V.I.; Lee, H.; Xu, F.; Yakobson, B.I. Carbyne from first principles: chain of C atoms, a nanorod or a nanorope. *ACS Nano* **2013**, 7, 10075-10082.

(8) Zanolli, Z.; Onida, G.; Charlier, J.C. Quantum spin transport in carbon chains. *ACS Nano* **2010**, 4, 5174-5180

(9) Bonardi, P.; Achilli, S.; Tantardini, G.F.; Martinazzo, R. Electron transport in carbon wires in contact with Ag electrodes: a detailed first principles investigation. *Phys. Chem. Chem. Phys.* **2015**, 17, 18413-18425

(10) Smogunov, A.; Dappe, Y.J. Symmetry-derived half metallicity in atomic and molecular junction. *Nano Lett.* **2015**, 15, 3552−3556

(11) Hsu, B.C.; Yao, H.T.; Liu, W.L.; Chen, Y.C. Oscillatory and sign-alternating behaviours of the Seebeck coefficients in carbon monoatomic junctions. *Phys. Rev. B* **2013**, 88, 115429

(12) Wang, M.; Lin, S. Ballistic thermal transport in carbyne and cumulene with micron-scale spectral acoustic phonon mean free path. *Scientific Rep.* **2015,** 5, 18122

(13) Zhu, Y.; Bai, H.; Huang, Y. Electronic Property Modulation of One-Dimensional Extended Graphdiyne Nanowires from a First-Principle Crystal Orbital View. *Chemistry Open* **2016**, 5, 78-87.

(14) Agarwal, N.R; Lucotti, A.; Tommasini, M.; Chalifoux, W.A.; Tykwinski, R.R. Nonlinear optical properties of polyynes: an experimental prediction for carbyne. *J. Phys. Chem. C* **2016**, 120, 11131-11139.

(15) Sorokin, P.B.; Lee, H.; Antipina, L.Y.; Singh, A.K.; Yakobson, B.I. Calcium-decortated carbyne networks as hydrogen storage media. *Nano Letters* **2011**, 11, 2660–2665.

(16) Cretu, O.; Botello-Mendez, A.R.; Janowska I.; Cuong, P.H.; Charlier, J.C.; Banhart, F. Electrical transport measured in atomic carbon chains. *Nano Letters* **2013,** 13, 3487−3493

(17) Standley, B.; Bao, W.; Zhang, H.; Bruck, J.; Lau, C.N.; Bockrath, M. Graphene-based atomic-scale switches. *Nano Lett.* **2008**, 8, 3345-3349.
(18) Milani, A.; Tommasini, M.; Russo, V.; Li Bassi, A.; Lucotti, A.; Cataldo, F.; Casari, C.S. Raman spectroscopy as a tool to investigate the structure and electronic properties of carbon-atom wires. *Beilstein J. Nanotechnol.*. **2015**, 6, 480-491.
(19) Ravagnan, L.; Siviero, F.; Lenardi, C.; Piseri, P.; Barborini, E.; Milani, P.; Casari, C.S.; Li Bassi, A.; Bottani, C.E. Cluster-beam deposition and in situ characterization of carbyne-rich carbon films. *Phys. Rev. Lett.* **2002**, 89, 285506
(20) Casari, C.S.; Li Bassi, A.; Baserga, A.; Ravagnan, L.; Piseri, P.; Lenardi, C.; Tommasini, M.; Milani, A.; Fazzi, D.; Bottani, C.E.; Milani, P. Low-frequency modes in the Raman spectrum of sp-sp(2) nanostructures carbon. *Phys. Rev. B* **2008**, 77, 195444
(21) Ferrari, A.C; Basko, D.M. Raman spectroscopy as a versatile tool for studying the properties of graphene. *Nature Nanotechnology* **2013**, 8, 235–246
(22) Dresselhaus, M.S.; Jorio A.; Dresselhaus, G.; Saito, R. Perspectives on carbon nanotubes and graphene Raman spectroscopy. *Nano Lett*. **2010**, 10, 751–758
(23) Chalifoux, W.A.; Tykwinski, R.R. Synthesis of polyynes to model the sp-carbon allotrope carbyne. *Nature Chemistry* **2010,** 2, 967-971
(24) Januszewski, J.A; Tykwinski, R.R. Synthesis and properties of long [n] cumulenes (n>=5). *Chem. Soc. Rev.*, **2014**, 43, 3184-3203
(25) Shi, L.; Rohringer, P.; Suenaga, K.; Niimi, Y.; Kotakoski J. et al. Confined linear carbon chains as a route to bulk carbyne. *Nature Materials* **2016**, 15, 634
(26) Casari, C.S; Giannuzzi, C.S.; Russo, V. Carbon-atom wires produced by nanosecond pulsed laser deposition in background gas. *Carbon* **2016**, 104, 190-195
(27) Polyynes: Synthesis Properties and Applications, F. Cataldo Ed., Taylor & Francis **2006**.
(28) Milani, A.; Lucotti, A.; Russo, V.; Tommasini, M.; Cataldo, F.; Li Bassi, A.; Casari, C.S. Charge Transfer and vibrational structure of sp-hybridized carbon atomic wires probed by surface enhanced Raman spectroscopy. *J. Phys. Chem. C* **2011**, 115, 12836–12843
(29) Bettini, L.G.; Della Foglia, F.; Piseri, P.; Milani, P. Interfacial properties of a carbyne-rich nanostructured carbon thin film in ionic liquid. *Nanotechnology* **2016**, 27 115403
(30) Kang, C.S.; Fujisawa, K.; Ko Y.I. et al. Linear carbon chains inside multi-walled carbon nanotubes: growth mechanism, thermal stability and electrical properties. *Carbon* **2016**, 107, 217- 224
(31) Jin, C.; Lan, H.P; Peng, L.M.; Suenaga, K; Iijima, S. Deriving carbon atomic chains from graphene. *Phys. Rev. Lett*. **2009,** 102, 205501.

(32) La Torre, A.; Botello-Mendez, A.; Baaziz, W.; Charlier, J.C.; Banhart, F. Strain-induced metal-semiconductor transition observed in atomic carbon chains. *Nature Comm*. **2015**, 6, 6636
(33) Sun, Q.; Cai, L.L.; Ma, H.H.; Yuan, C.X.; Xu, W. Dehalogenative homocoupling of terminal alkynyl bromides on Au(111): Incorporation of acetylenic scaffolding into surface nanostructures. *ACS Nano*, **2016**, 10, 7023–7030
(34) Ivanovskii, A.L Graphynes and graphdyines. *Progr. Solid State Chem*. **2013**, 41, 1-19.
(35) Malko, D.; Neiss, C.; Vines, F.; Görling, A. Competition for graphene: graphynes with direction-dependent Dirac cones. *Phys. Rev. Lett.* **2012**, 108, 086804
(36) A. Lucotti, M. Tommasini, M. Del Zoppo, C. Castiglioni, G. Zerbi, F. Cataldo, C.S. Casari, A. Li Bassi, V. Russo, M. Bogana, C.E. Bottani Raman and SERS investigation of isolated sp carbon chains. *Chemical Physics Letters* **2006**, 417, 78-82
(37) A. Lucotti, C. S. Casari, M. Tommasini, A. Li Bassi, D. Fazzi, V. Russo, M. Del Zoppo, C. Castiglioni, F. Cataldo, C. E. Bottani, G. Zerbi sp carbon chain interaction with silver nanoparticles probed by Surface Enhanced Raman Scattering. *Chemical Physics Letters* **2009**, 478, 45–50
(38) Tabata, H.; Fujii,M.; Hayashi, S.; Doi, T.; Wakabayashi,T.; *Carbon* **2006**, 44, 3168–3176.
(39) Cataldo, F.; Ravagnan, L.; Cinquanta, E.; Castelli, I.E.; Manini, N.; Onida, G.; Milani, P. Synthesis, characterization and modeling of naphtyl-terminated sp carbon chains: Dinaphtylpolyynes. *J. Phys. Chem. B* **2010**, 114, 14834-14841.
(40) Cinquanta, E.; Ravagnan, L.; Castelli, I.E.; Cataldo, F.; Manini, N.; Onida, G.; Milani, P. Vibrational characterization of dinaphtylpolyynes: a model system for the study of end-capped sp carbon chains. *J. Chem. Phys.* **2011**, 135, 194501.
(41) Fazzi, D.; Scotognella, F.; Milani, A.; Brida, D.; Manzoni, C.; Cinquanta, E.; Devetta, M.; Ravagnan, L.; Milani, P.; Cataldo, F.; Luer, L.; Wannemacher, R.; Cabanillas-Gonzales, J.; Negro, M.; Stagira, S.; Vozzi, C. Ultrafast spectroscopy of linear carbon chains: the case of dinaphtylpolyynes. *Phys. Chem. Chem. Phys*. **2013**, 15, 9384-9391.
(42) Rivelino, R.; dos Santos, R.B.; Mota, F.D.; Gueorguiev, G.K. Conformational effects on structure, electron states, and Raman scattering properties of linear carbon chains terminated by graphene-like pieces. *J. Phys. Chem. C* **2010**, 114, 16367-16372.
(43) Januszewski, J.A.; Wendinger, D.; Methfessel, C.D.; Hampel, F.; Tykwinski, R.R. Synthesis and structure of tetraaerylcumulenes: characterization and bond-length alternation versus molecular length. *Angew. Chem. Int. Ed.* **2013**, 52, 1817-1821

(44) Tommasini, M.; Milani, A.; Fazzi, D.; Lucotti, A.; Castiglioni, C.; Januszewski, J.A.; Wendinger, D.; Tykwinski, R.R. pi-conjugation and endgroup effects in long cumulenes: Raman spectroscopy and DFT calculations. *J. Phys. Chem. C* **2014**, 118, 26415-26425

(45) Castiglioni, C.; Tommasini, M.; Zerbi, G. Raman spectroscopy of polyconjugated molecules and materials: confinement effect in one and two dimensions. *Phil. Trans. Roy. Soc. A* **2004**, 362, 2425-2459.

(46) Tommasini, M.; Fazzi, D.; Milani, A.; Del Zoppo, M.; Castiglioni, C.; Zerbi, G. Intramolecular vibrational force fields for linear carbon chains through an adaptative linear scaling scheme. *J. Phys. Chem. A* **2007**, 111, 11645-11651.

(47) Milani, A.; Tommasini, M.; Del Zoppo, M.; Castiglioni, C.; Zerbi, G. Carbon nanowires: phonon and pi-electron confinement. *Phys. Rev. B* **2006**, 74, 153418

(48) Milani, A.; Tommasini, M.; Zerbi, G. Connection among Raman wavenumbers, bond length alternation and energy gap in polyynes. *J. Raman Spectrosc*. **2009**, 40, 1931-1934.

(49) Cataldo, F. Synthesis of polyynes in a submerged electric arc in organic solvents. *Carbon* **2004**, 42, 129-142

(50) Chalifoux, W.A; McDonald, R.; Ferguson, M.J.; Tykwinski, R.R. tert-Butyl-end-capped polyynes: crystallographic evidence of reduced bond-length alternation. *Angew. Chemie Int. Ed.* **2009**, 48, 7915-7919

(51) Innocenti, F.; Milani, A.; Castiglioni, C. Can Raman spectroscopy detect cumulenic structures of linear carbon chains? *J. Raman Spectrosc*. **2010**, 41, 226-236.

(52) Cataldo, F. Evidences about carbyne formation together with other carbonaceous materials by thermal decomposition of diiodoacetylene. *Fullerenes Nanot. Carbon Nanostruct*. **2001**, 9, 525-542

(53) Adamo, C.; Barone, V. Toward reliable density functional methods without adjustable parameters: the PBE0 model. *J. Chem. Phys.* **1999**, 110, 6158-6170.

(54) Gaussian 09, Revision E.01, Frisch, M. J.; Trucks, G. W.; Schlegel, H. B.; Scuseria, G. E.; Robb, M. A.; Cheeseman, J. R.; Scalmani, G.; Barone, V.; Mennucci, B.; Petersson, G. A.; Nakatsuji, H.; Caricato, M.; Li, X.; Hratchian, H. P.; Izmaylov, A. F.; Bloino, J.; Zheng, G.; Sonnenberg, J. L.; Hada, M.; Ehara, M.; Toyota, K.; Fukuda, R.; Hasegawa, J.; Ishida, M.; Nakajima, T.; Honda, Y.; Kitao, O.; Nakai, H.; Vreven, T.; Montgomery, J. A., Jr.; Peralta, J. E.; Ogliaro, F.; Bearpark, M.; Heyd, J. J.; Brothers, E.; Kudin, K. N.; Staroverov, V. N.; Kobayashi, R.; Normand, J.; Raghavachari, K.; Rendell, A.; Burant, J. C.; Iyengar, S. S.; Tomasi, J.; Cossi, M.; Rega, N.; Millam, J. M.; Klene, M.; Knox, J. E.; Cross, J. B.; Bakken, V.; Adamo, C.; Jaramillo, J.; Gomperts, R.; Stratmann, R. E.; Yazyev, O.; Austin, A. J.; Cammi, R.; Pomelli, C.; Ochterski, J. W.; Martin, R. L.; Morokuma, K.; Zakrzewski, V. G.; Voth,

G. A.; Salvador, P.; Dannenberg, J. J.; Dapprich, S.; Daniels, A. D.; Farkas, Ö.; Foresman, J. B.; Ortiz, J. V.; Cioslowski, J.; Fox, D. J. Gaussian, Inc., Wallingford CT, **2009**.
(55) Tommasini, M.; Fazzi, D.; Milani, A.; Del Zoppo, M.; Castiglioni, C.; Zerbi, G. Intramolecular vibrational force fields for linear carbon chains through an adaptative linear scaling scheme. *J. Phys. Chem. A* **2007**, 111, 11645-11651.
(56) Milani, A.; Tommasini, M.; Zerbi, G. Carbynes phonons: a tight binding force field. *J. Chem. Phys.* **2008**, 128, 064501.
(57) Yang, S.J.; Kertesz, M.; Zolyomi, V.; Kurti, J. Application of a novel linear/exponential hybrid force field scaling scheme to the longitudinal Raman active mode of polyyne. *J. Phys. Chem. A* **2007**, 111, 2434-2441